\documentclass[a4paper,10pt,oneside,onecolumn,
nonatbib]{elsarticle}
\usepackage[english]{babel}
\usepackage{graphicx}
\usepackage[breaklinks=true,colorlinks=true,linkcolor=blue,urlcolor=blue,citecolor=blue]{hyperref}
\usepackage{lipsum}
\usepackage{multirow}
\usepackage[tableposition=top]{caption}

\usepackage{csquotes}
\usepackage{algorithm}
\usepackage{algpseudocode}
\usepackage{tikz}
\usetikzlibrary{positioning, shapes.multipart, fit, arrows.meta}
\usepackage{pgfplots}
\pgfplotsset{compat=1.17}
\usepackage[acronym]{glossaries-extra}
\setabbreviationstyle[acronym]{long-short}
\glsdisablehyper
\usepackage[all]{nowidow}
\usepackage{siunitx}
\DeclareSIUnit\clight{\text{\ensuremath{c}}}
\DeclareSIUnit\mhz{\mega\hertz}
\DeclareSIUnit\khz{\kilo\hertz}
\DeclareSIUnit\mm{\milli\metre}
\DeclareSIUnit\cm{\centi\metre}
\DeclareSIUnit\ms{\milli\second}
\DeclareSIUnit\um{\micro\metre}
\DeclareSIUnit\ns{\nano\second}
\DeclareSIUnit\mev{\mega\electronvolt}
\DeclareSIUnit\mevc{\mega\electronvolt\per\clight}
\DeclareSIUnit\mevsc{\mega\electronvolt\per\clight\squared}
\DeclareSIUnit\kev{\kilo\electronvolt}
\DeclareSIUnit\kevc{\kilo\electronvolt\per\clight}
\DeclareSIUnit\kevsc{\kilo\electronvolt\per\clight\squared}

\newacronym{sificc}{SiFi\babelhyphen{nobreak}CC}{\textbf{Si}licon Photomultiplier and Scintillating \textbf{Fi}ber based \textbf{C}ompton \textbf{C}amera}
\newacronym{sipm}{SiPM}{silicon photomultiplier}
\newacronym{pde}{PDE}{photon detection efficiency}
\newacronym{pbs}{PBS}{pencil beam scanning}
\newacronym{rbe}{RBE}{relative biological effectiveness}
\newacronym{PMMA}{PMMA}{polymethyl methacrylate}
\newacronym[shortplural={PGs},longplural=prompt gammas]{PG}{PG}{prompt gamma}
\newacronym{PGT}{PGT}{prompt gamma timing}
\newacronym{geant}{Geant4}{GEometry ANd Tracking}
\newacronym{LUT}{LUT}{LookUpTable}
\newacronym{goddess}{GODDeSS}{\textbf{G}eant4 \textbf{O}bjects for \textbf{D}etailed \textbf{De}tectors with \textbf{S}cintillators and \textbf{S}iPMs}
\newacronym{mlem}{LM\babelhyphen{nobreak}MLEM}{list-mode maximum-likelihood expectation maximisation}
\newacronym{fpga}{FPGA}{Field Programmable Gate Array}
\newacronym{tdc}{TDC}{time to digital converter}
\newacronym{pcb}{PCB}{printed circuit board}
\newacronym{fee}{FEE}{front-end electronics}
\newacronym{daq}{DAQ}{data acquisition}
\newacronym[shortplural={GAs},longplural=genetic algorithms]{ga}{GA}{genetic algorithm}
\newacronym{sad}{SAD}{scatterer-absorber distance}
\newacronym{ssd}{SSD}{source-scatterer distance}
\newacronym{sl}{SL}{scatterer layers}
\newacronym{al}{AL}{absorber layers}
\newacronym{pla}{PLA}{polylactic acid}
\newacronym{OPM}{OPM}{optical photon model}
\newacronym{feedaq}{FEE+DAQ}{front-end electronics and data acquisition}
\newacronym{llr}{LLR}{low-level reconstruction}
\newacronym{es}{ES}{event selection}
\newacronym{dca}{DCA}{distance of closest approach}
\newacronym{ir}{IR}{image reconstruction}
\newacronym{ff}{FF}{first fibre}
\newacronym{hf}{HF}{heaviest fibre}
\newacronym{ap}{AP}{averaged position}

\makeatletter
\let\c@author\relax
\makeatother
\usepackage[
backend=biber,
style=nejm,
citestyle=numeric-comp,
articledoi=true,
url=true,
]{biblatex}
\usepackage{tabularx}
\usepackage{booktabs}
\usepackage[capitalise]{cleveref}

\usepackage{setspace}

\graphicspath{{figures/}}

\begin{document}
\begin{frontmatter}
\title{Genetic algorithm as a tool for detection setup optimisation 
for prompt-gamma imaging in proton therapy:
SiFi Compton camera case study}

\author[a1]{Jonas~Kasper}
\author[a1,a2]{Awal~Awal}
\author[a1]{Ronja~Hetzel\fnmark[fn1] }
\author[a3,a4]{Magdalena~Ko{\l}odziej\fnmark[fn3]}
\author[a3]{Katarzyna~Rusiecka}
\author[a1]{Achim~Stahl}
\author[a3]{Ming-Liang~Wong\fnmark[fn2] }
\author[a3]{Aleksandra~Wro\'nska\corref{cor1}}\ead{aleksandra.wronska@uj.edu.pl}

\cortext[cor1]{Corresponding author}

\fntext[fn1]{Present address: Biophysics Department, GSI Helmholtzzentrum für Schwerionenforschung GmbH, Darmstadt, Germany.}

\fntext[fn3]{Present address: Institute of Medical Engineering, University of Lübeck, Lübeck, Germany.}

\fntext[fn2]{Present address: Department of Physics, University of Liverpool, Liverpool, UK.}

\address[a1]{III. Physikalisches Institut B, RWTH Aachen University, Aachen, Germany}
\address[a2]{Forschungszentrum J{\"u}lich, J{\"u}lich, Germany}
\address[a3]{Marian Smoluchowski Institute of Physics, Jagiellonian University, Krak\'ow, Poland}
\address[a4]{Doctoral School of Exact and Natural Sciences, Jagiellonian University, Kraków, Poland}

\begin{abstract}
\textbf{Objective:} Proton therapy is a precision-focused cancer treatment where accurate proton beam range monitoring is critical to ensure effective dose delivery. This can be achieved by prompt gamma detection with a Compton camera like the SiFi-CC. This study aims to show the feasibility of optimising the geometry of SiFi-CC Compton camera for verification of dose distribution via prompt gamma detection a genetic algorithm (GA). 
\\
\textbf{Approach:} The SiFi-CC key geometric parameters for optimisation with the GA are the source-to-scatterer and scatterer-to-absorber distances, and the module thicknesses. 
The optimisation process was conducted with a software framework based on the Geant4 toolkit, which included detailed and realistic modelling of gamma interactions, detector response, and further steps such as event selection and image reconstruction. The performance of each individual configuration was evaluated using a fitness function incorporating factors related to gamma detection efficiency and image resolution.
\\
\textbf{Results:} The GA-optimised SiFi-CC configuration demonstrated the capability to detect a \SI{5}{\mm} proton beam range shift with a \SI{2}{\mm} resolution using \num{5e8} protons. The best-performing geometry, with 16 fibre layers in the scatterer, 36 layers in the absorber, source-to-scatterer distance \SI{150}{\mm} and scatterer-to-absorber distance \SI{120}{\mm}, has an imaging sensitivity of \num{5.58(1)e-5}.
\\
\textbf{Significance:} This study demonstrates that the SiFi-CC setup, optimised through a GA, can reliably detect clinically relevant proton beam range shifts, improving real-time range verification accuracy in proton therapy. The presented implementation of a GA is a systematic and feasible way of searching for a SiFi-CC geometry that shows the best performance.
\end{abstract}

\end{frontmatter}

\section{Introduction}
Online monitoring in proton therapy is one of the most researched topics of the last decade in the community. Different approaches are in research or have already started the clinical evaluation phase \cite{Richter2016,Berthold2023,Xie2017,Xie2020,Gonzalez2018}. Most of them utilise by-products of the proton irradiation of the patient, many projects focus on the detection of \gls{PG} radiation, reviewed e.g. in~\cite{Wronska2021,Pinto2024}. Among these, the concept of a Compton camera is particularly interesting, as it provides possible access to a three-dimensional dose distribution. However, the realisation of this concept for online beam range verification in the clinical environment is very demanding both software- and hardware-wise~\cite{Pausch2018,Polf2021,Golnik2016,Liprandi2017,KrimmerClarys2017}.
The Compton camera operates based on the principle of Compton scattering, where an incoming gamma photon undergoes scattering in the first detection module, known as the scatterer, followed by absorption of the scattered photon in a second module, referred to as the absorber. By registering coincident interactions in both modules and determining the associated positions and energy deposits, the direction of the primary gamma photon can be constrained to a conical surface. Image reconstruction algorithms utilise the collection of such cones to estimate the spatial distribution of the prompt gamma emission origins. 

In proton therapy monitoring, the optimisation of \gls{PG}-based setups is necessary and commonly carried out using Monte Carlo simulations, see e.g.~\cite{Cambraia2012,Pinto2014,Morozov2021}. However, this presents a multi-parameter challenge, with variables such as the thickness and relative positioning of the two modules in a Compton camera, or the collimator thickness, layout, and detector pixel size for gamma cameras, whether equipped with a single slit or multiple parallel slits. The realistic simulation of setup responses, event selection, followed by image reconstruction, is highly CPU-intensive, making a classical full scan of the parameter space impractical. Therefore, alternative approaches to setup optimisation must be explored.

\begin{figure}[!htp]
    \centering
    \includegraphics[width=0.8\linewidth]{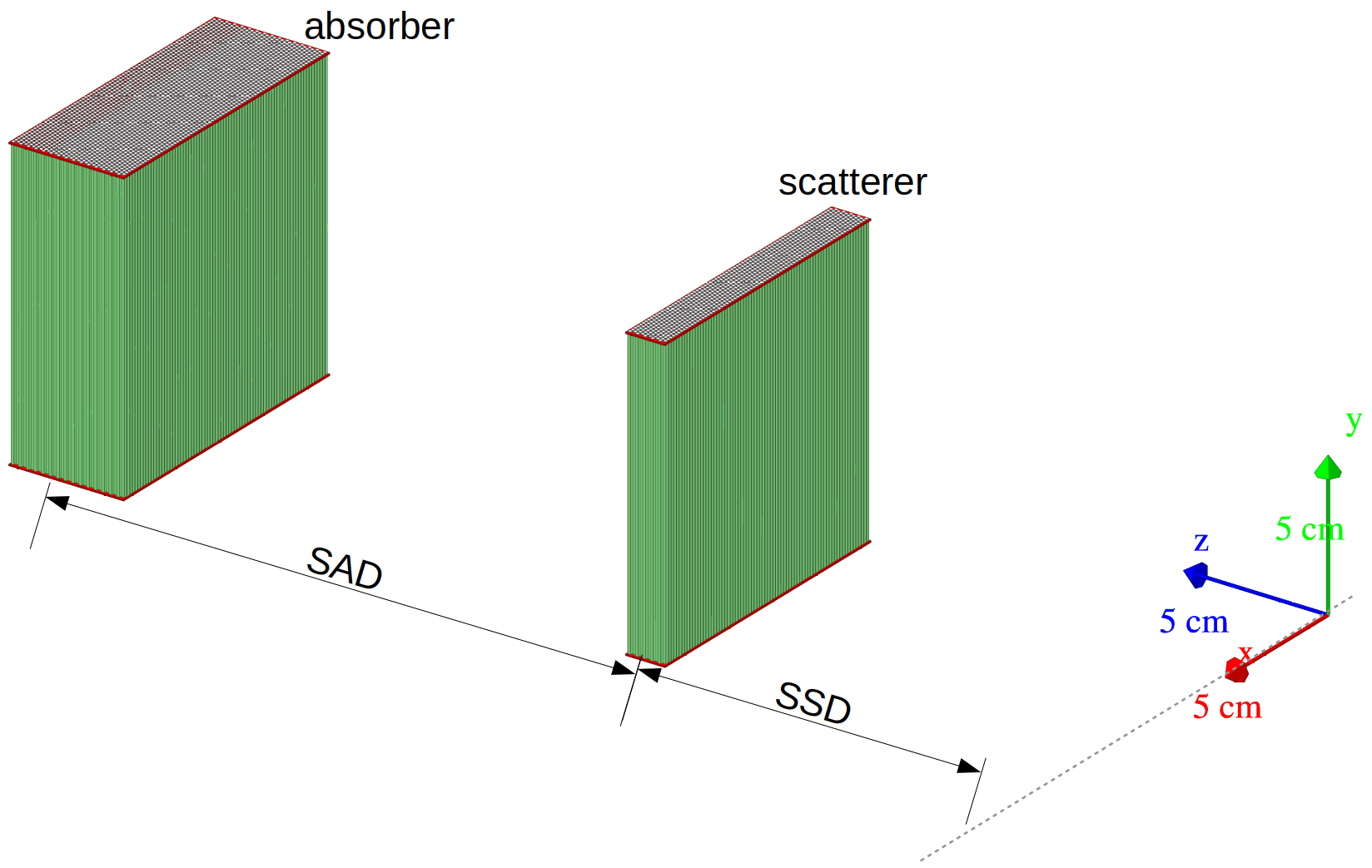}
    \caption{Perspective view of an example Compton camera: the \gls{sificc} setup, with the absorber and the scatterer modules built with scintillating LYSO:Ce fibres (green) read out by silicon photomultipliers (grey). The source is assumed to be located in the $XY$ plane. The indicated distances (SAD, SSD) and module thicknesses are to be optimised.}
    \label{figure:SiFiCCSketch}
\end{figure}

This study, as part of the broader R\&D presented in detail in~\cite{KasperPhD}, focuses on the \gls{sificc} as an example of a Compton camera. As shown in \cref{figure:SiFiCCSketch}, each module of this setup consists of layers of $1\times1\times 100$~mm$^3$ LYSO:Ce scintillating fibres read out by \glspl{sipm} at both ends~\cite{Kasper2020}.
\section{Methods}

\subsection{Genetic algorithm}
Among the optimisation tools, \glsxtrfullpl{ga} belong to metaheuristic algorithms, and are inspired by the process of biological evolution~\cite{Holland1975}. A candidate solution of the posed optimisation problem is called an \emph{individual}. It is characterised by a set of \emph{genes}, i.e. a unique set of values of parameters to be optimised. Genetic algorithms operate on \emph{generations} or \emph{populations} that are formed by a fixed number of individuals. 
The workflow of a \gls{ga} 
starts with the first generation, composed of $N$ randomly generated individuals. The fitness of each of them, i.e. a measure of how well an individual solves the posed problem, is evaluated by means of a fitness function, which must be defined for each problem individually.

Based on the obtained fitness values, individuals that pass their genes to the next generation are selected. The following step is called crossover and it defines a way the new generation is formed from the previous one. Finally, the mutation step randomly re-rolls some of the genes within the population to enable exploration of the parameter space.
The presented cycle is repeated until a stop criterion is met.

\glspl{ga} applied to problems being subject to statistical precision, e.g. Monte Carlo simulations, also suffer from statistical fluctuations. This means that the same individual, when assessed twice, can yield different values of fitness. Other known weaknesses of \glspl{ga} are lack of convergence due to inappropriate choice of applied operators, or premature convergence to a local maximum, resulting in finding not the optimal, but a reasonably good solution~\cite{Beasley1993,Yang2021Ch6}.

The workflow of the \gls{ga} used for the \gls{sificc} setup optimisation problem is depicted in~\cref{algo:ga}. The following sections describe its settings and conditions in more detail.

\begin{algorithm}
\begin{algorithmic}[1]
\State Initialise population with $N$ random individuals
\While {not converged or not 10th generation}
    \State \textit{Evaluate} fitness of offspring
    \State \textit{Select} top 80\% from the population based on fitness
    \State Apply gene pool method for \textit{crossover} from parents to generate offspring
    \State Apply \textit{mutation} to some offspring
\EndWhile
\State \textbf{return} best solution found
\end{algorithmic}
\caption{The algorithm converges when the fitness of three consecutive generations differ by less than 5\%.}
\label{algo:ga}
\end{algorithm}

\subsubsection{Optimisation targets}
Within this framework,
four strongly correlated parameters are tuned, the two distances \gls{ssd} and \gls{sad} and the numbers of layers in the scatterer (\gls{sl}) and in the absorber (\gls{al}). For both distances \gls{ssd} and \gls{sad} the parameter space is restricted to the range between \SI{90}{\mm} and \SI{300}{\mm}, with a \SI{30}{\mm} step size. The upper limit is imposed to make sure that the setup maintains a reasonable acceptance for the envisaged application, i.e. a too low coincidence rate is avoided. The numbers of layers in the modules increase in steps of 2, whereby \gls{sl} takes values between 2 and 20 and \gls{al} between 20 and 40 layers. The limits are driven by the goal to reach a high gamma efficiency while still keeping a reasonable number of readout channels.
  The simulation parameters defining an individual, their allowed variation ranges and step sizes, are given in~\cref{table:individualpars}.
\begin{table}[!htb]
\centering
\begin{tabular}{l|l|l|l}
Parameter	&Lower limit&Upper limit&Step size\\
\hline
\gls{ssd}&\SI{90}{\mm}&\SI{300}{\mm}&\SI{30}{\mm}\\
\gls{sad}&\SI{90}{\mm}&\SI{300}{\mm}&\SI{30}{\mm}\\
\gls{sl}&2&20&2\\
\gls{al}&20&40&2\\
\end{tabular}
\caption{Parameter space describing possible individuals in the \gls{ga}. The lower limits on distances were determined by physical constraints such as holding structures, etc., while the upper limits are to prevent acceptance reduction and thus a too low coincidence rate.\label{table:individualpars}}
\end{table}

To minimise the optimisation load on the \gls{ga}, parameters other than listed above, that were found not to be directly correlated with them, were tuned one-by-one before, and used at their optimal, fixed values in the course of the \gls{ga}. On the geometrical side, the number of fibres forming a single layer was fixed at 76. Other simulation and setup parameters are discussed in \cref{sec:evaluation}.

\subsubsection{Fitness function}
The fitness function $f(i)$ is a crucial tool to evaluate and rank the performance of individuals and must be built so that all aspects of performance, which are to be optimised, are taken into account. The fitness function has been constructed so that the different aspects that describe the fitness of an individual are separated as much as possible. Evaluation of a single individual requires the execution of the full simulation and analysis chain discussed in \cref{sec:evaluation}. 

For a fixed number of shot protons $N_\mathrm{Protons}$, the geometric changes in the setup lead to different numbers of prompt gamma that cause a coincidence in the detector suitable for a Compton cone reconstruction, also called distributed Compton events $N_\mathrm{Dist.Comp.}$. Larger numbers of events available in reconstruction are favourable for a Compton camera, since they allow to suppress statistical fluctuations in the reconstructed image. Consequently, the ratio between the numbers of signal events and impinging protons builds the first part of the fitness function, see \cref{equation:fitnessfunction}. 

The setup geometry has a similar influence on the number of recorded background events $N_\mathrm{Back.}$, i.e. random coincidences and other interactions besides the Compton effect leading to coincidences. The better the ratio between signal events and background events, the better the performance of the setup, hence the second factor in~\cref{equation:fitnessfunction}.

This original signal-to-background ratio is modified by \gls{es}. The quality of selection will be different for different setups; this builds the third part of the fitness function. Here, $N_\mathrm{Sel.Dist.Comp.}$ stands for the number of all distributed Compton events that are correctly selected, and $N_\mathrm{Sel.Back.}$ denotes the number of all background events that are falsely selected as Compton events.

The last part of the fitness function is given by the so-called clean image resolution $\sigma_\mathrm{CleanImg.}$, and it aims to describe the fitness of an individual in terms of resolution in the determination of the distal fall-off position.
To derive it, we reconstruct the images multiple times, based on fixed-size samples  of events, created using boostrapping on the whole sample of correctly identified distributed Compton events for a given individual. Standard deviation of the determined distal fall-off positions is the value of $\sigma_\mathrm{CleanImg.}$. More details on the procedure are given in \cref{sec:evaluation}.
  The clean image resolution is in the first order free from the influence of changes in acceptance due to the changes in \gls{ssd} and \gls{sad} because fixed-size samples are taken for image reconstruction, as well as from the event selection, since only correctly identified events are considered. Those effects are represented by other factors in the formula. The factor $\sigma_\mathrm{CleanImg.}$ is mostly sensitive to different sampling of the cone phase space available for different individuals, and the different position- and energy resolutions for different cone phase space regions. 
 
 Thus, the final fitness function is given by:
\begin{equation}
\label{equation:fitnessfunction}
f(i)=\frac{N_\mathrm{Dist.Comp.}}{N_{\mathrm{Protons}}}\frac{N_\mathrm{Dist.Comp.}}{N_\mathrm{Back.}}\frac{N_ \mathrm{Sel.Dist.Comp.}}{N_\mathrm{Sel.Back.}}\frac{1}{\sigma_\mathrm{CleanImg.}}
\end{equation}

\subsubsection{Convergence criterion}
Two basic parameters must be set that describe the algorithm: the number of individuals per generation and the number of generations evaluated. The limitation in computing power and the time consumption for assessment of a single individual performance lead to a maximum of ten generations that are considered in this approach, where each generation holds ten different individuals. If the generation fitness, defined as the sum of all fitness values in a generation, converges before, the algorithm is stopped. The convergence criterion used requires that three consecutive generations differ in their sums of fitness values not more than \SI{5}{\percent}. 

\subsubsection{Selection}
\begin{figure}[!htb]
    \centering
    \begin{tikzpicture}
    
    \definecolor{green}{rgb}{0.56, 0.93, 0.56}
    \definecolor{junglegreen}{rgb}{0.16, 0.67, 0.53}
    \definecolor{yellow}{rgb}{1.0, 1.0, 0.6}
    \definecolor{red}{rgb}{1.0, 0.4, 0.4}
    \definecolor{babyblue}{rgb}{0.54, 0.81, 0.94}

    \draw [decorate,decoration={brace,amplitude=9pt},thick]
        (-1.1,3.45) -- node[left=35pt,rotate=0,anchor=center,align=center] {1.666 which \\ is 80\% of \\ total fitness} (-1.1,5.45);

    \node at (-0.5, 5.25) {0.706};
    \node at (-0.5, 4.75) {0.392};
    \node at (-0.5, 4.25) {0.333};
    \node at (-0.5, 3.75) {0.235};
    \node at (-0.5, 3.25) {0.196};
    \node at (-0.5, 2.75) {0.078};
    \node at (-0.5, 2.25) {0.059};
    \node at (-0.5, 1.75) {0.041};
    \node at (-0.5, 1.25) {0.020};
    \node at (-0.5, 0.75) {0.010};
    
    \draw[fill=green] (0,5) rectangle (1,5.5) node[pos=.5] {A};
    \draw[fill=yellow] (0,4.5) rectangle (1,5) node[pos=.5] {B};
    \draw[fill=red] (0,4) rectangle (1,4.5) node[pos=.5] {C};
    \draw[fill=babyblue] (0,3.5) rectangle (1,4) node[pos=.5] {D};
    \draw[fill=white] (0,3) rectangle (1,3.5) node[pos=.5] {E};
    \draw[fill=white] (0,2.5) rectangle (1,3) node[pos=.5] {F};
    \draw[fill=white] (0,2) rectangle (1,2.5) node[pos=.5] {G};
    \draw[fill=white] (0,1.5) rectangle (1,2) node[pos=.5] {H};
    \draw[fill=white] (0,1) rectangle (1,1.5) node[pos=.5] {I};
    \draw[fill=white] (0,0.5) rectangle (1,1) node[pos=.5] {J};
    
    \draw[->, thick] (1.5, 3) -- (2, 3);
    
    \draw[fill=green] (2.7,5) rectangle (3.7,5.5) node[pos=.5] {A};
    \draw[fill=green] (2.7,4.5) rectangle (3.7,5) node[pos=.5] {A};
    \draw[fill=green] (2.7,4) rectangle (3.7,4.5) node[pos=.5] {A};
    \draw[fill=green] (2.7,3.5) rectangle (3.7,4) node[pos=.5] {A};
    \draw[fill=yellow] (2.7,3) rectangle (3.7,3.5) node[pos=.5] {B};
    \draw[fill=yellow] (2.7,2.5) rectangle (3.7,3) node[pos=.5] {B};
    \draw[fill=red] (2.7,2) rectangle (3.7,2.5) node[pos=.5] {C};
    \draw[fill=red] (2.7,1.5) rectangle (3.7,2) node[pos=.5] {C};
    \draw[fill=babyblue] (2.7,1) rectangle (3.7,1.5) node[pos=.5] {D};
    \draw[fill=babyblue] (2.7,0.5) rectangle (3.7,1) node[pos=.5] {D};
    
    \end{tikzpicture}
    \caption{For this selection example, the individuals (in descending fitness score) up to D contribute 80\% to the total generation fitness (1.666 out of 2.07). These four individuals are selected with abundance corresponding to their fitness. Adapted from ~\cite{Meloni2016}.}
    \label{figure:selectionmechanism}
\end{figure}
The selection mechanism chosen in this \gls{ga} is inspired by~\cite{Meloni2016} and follows a concept that is a combination of the roulette wheel selection, the rank selection, and the principle of elitism. Starting from the individual with the highest fitness, the individuals building the top \SI{80}{\percent} of the generation fitness are selected to give their genes to the next generation. The total of ten individuals are recreated by reproducing the chosen individuals with the abundance equal to the ratio of their fitness to the sum of the fitness of all selected individuals as depicted in \cref{figure:selectionmechanism}. At the end of this stage, some individuals may show the same settings. Furthermore, the principle of elitism ensures that the individual with the highest fitness transfers unchanged to the next generation.

\subsubsection{Crossover}
\begin{figure}
    \centering
    \begin{tikzpicture}
    \node[rotate=90] at (-0.3,5.5) {Parents};
    \draw[fill=white] (0,6) rectangle (0.5,6.5) node[pos=.5] {A};
    \draw[fill=white] (0,5.5) rectangle (0.5,6) node[pos=.5] {B};
    \draw[fill=white] (0,5) rectangle (0.5,5.5) node[pos=.5] {C};
    \draw[fill=white] (0,4.5) rectangle (0.5,5) node[pos=.5] {D};
    
    \foreach \y in {4.5, 5, 5.5, 6} {
        \foreach \x in {0.5, 1, 1.5} {
            \draw (\x,\y) rectangle (\x+1,\y+0.5);
        }
    }
    \node at (0.75,6.25) {$a_1$}; \node at (1.25,6.25) {$a_2$}; \node at (1.75,6.25) {$a_3$}; \node at (2.25,6.25) {$b_4$};
    \node at (0.75,5.75) {$a_1$}; \node at (1.25,5.75) {$c_2$}; \node at (1.75,5.75) {$a_3$}; \node at (2.25,5.75) {$b_4$};
    \node at (0.75,5.25) {$a_1$}; \node at (1.25,5.25) {$b_2$}; \node at (1.75,5.25) {$a_3$}; \node at (2.25,5.25) {$b_4$};
    \node at (0.75,4.75) {$b_1$}; \node at (1.25,4.75) {$a_2$}; \node at (1.75,4.75) {$b_3$}; \node at (2.25,4.75) {$b_4$};
    
    \draw[->, thick] (2.7, 5.75) -- (3.2, 5.75);
    
    \node[rotate=90] at (3.5,5.5) {Gene pool};
    \draw (3.8,5.5) rectangle (4.3,6) node[pos=.5] {$a_1$}; \draw (4.3,5.5) rectangle (4.8,6) node[pos=.5] {$c_2$}; \draw (4.8,5.5) rectangle (5.3,6) node[pos=.5] {$a_3$}; \draw (5.3,5.5) rectangle (5.8,6) node[pos=.5] {$b_4$};
    \draw (3.8,5) rectangle (4.3,5.5) node[pos=.5] {$a_1$}; \draw (4.3,5) rectangle (4.8,5.5) node[pos=.5] {$b_2$}; \draw (4.8,5) rectangle (5.3,5.5) node[pos=.5] {$a_3$}; \draw (5.3,5) rectangle (5.8,5.5) node[pos=.5] {$b_4$};
    \draw (3.8,4.5) rectangle (4.3,5) node[pos=.5] {$b_1$}; \draw (4.3,4.5) rectangle (4.8,5) node[pos=.5] {$a_2$}; \draw (4.8,4.5) rectangle (5.3,5) node[pos=.5] {$b_3$}; \draw (5.3,4.5) rectangle (5.8,5) node[pos=.5] {$b_4$};

    \draw[->, thick] (5.9, 5.75) -- (6.4, 5.75);
    
    \node[rotate=90] at (6.7,5.5) {Shuffling};
    
    \draw[->, thick] (7, 5.75) -- (7.5, 5.75);
    
    \node[rotate=90] at (7.7,5.5) {Offspring};
    \draw[fill=white] (8,6) rectangle (8.5,6.5) node[pos=.5] {A};
    \draw[fill=white] (8,5.5) rectangle (8.5,6) node[pos=.5] {X};
    \draw[fill=white] (8,5) rectangle (8.5,5.5) node[pos=.5] {Y};
    \draw[fill=white] (8,4.5) rectangle (8.5,5) node[pos=.5] {Z};

    \foreach \y in {4.5, 5, 5.5, 6} {
        \foreach \x in {8.5, 9, 9.5} {
            \draw (\x,\y) rectangle (\x+1,\y+0.5);
        }
    }
    
    \node at (8.75,6.25) {$a_1$}; \node at (9.25,6.25) {$a_2$}; \node at (9.75,6.25) {$a_3$}; \node at (10.25,6.25) {$b_4$};
    \node at (8.75,5.75) {$a_1$}; \node at (9.25,5.75) {$b_2$}; \node at (9.75,5.75) {$a_3$}; \node at (10.25,5.75) {$b_4$};
    \node at (8.75,5.25) {$a_1$}; \node at (9.25,5.25) {$c_2$}; \node at (9.75,5.25) {$a_3$}; \node at (10.25,5.25) {$b_4$};
    \node at (8.75,4.75) {$a_1$}; \node at (9.25,4.75) {$a_2$}; \node at (9.75,4.75) {$b_3$}; \node at (10.25,4.75) {$b_4$};
    
    \end{tikzpicture}
    
    \caption{As an example, only four parents are considered for this crossover step. The genes of the fittest individual remain unchanged whereas the others will get their new genes from a pool. Adapted from ~\cite{Meloni2016}.}
    \label{figure:crossovermechanism}
\end{figure}
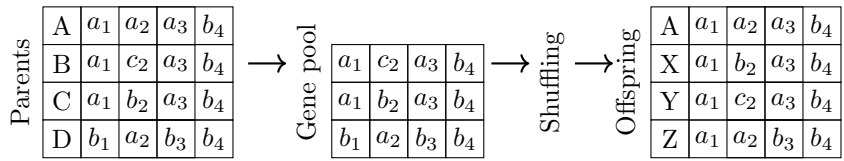

Due to the principle of elitism, the fittest individual of the mother generation does not change. The other nine individuals entering the new generation are created using the gene pool method. For each gene, a pool is generated from the values of the selected individuals, see \cref{figure:crossovermechanism}. Subsequently, the values are shuffled and randomly assigned to the individuals of the next generation~\cite{Muehlenbein1996}.  

\subsubsection{Mutation}
\begin{figure}
    \centering
    \begin{tikzpicture}
    \definecolor{green}{rgb}{0.56, 0.93, 0.56}
    \definecolor{babyblue}{rgb}{0.54, 0.81, 0.94}
    \draw[fill=white] (0,6) rectangle (0.5,6.5) node[pos=.5] {A};
    \draw[fill=white] (0,5.5) rectangle (0.5,6) node[pos=.5] {B};
    \draw[fill=white] (0,5) rectangle (0.5,5.5) node[pos=.5] {C};
    \draw[fill=white] (0,4.5) rectangle (0.5,5) node[pos=.5] {D};
    
    \foreach \y in {4.5, 5, 5.5, 6} {
        \foreach \x in {0.5, 1, 1.5} {
            \draw (\x,\y) rectangle (\x+1,\y+0.5);
        }
    }
    \node at (0.75,6.25) {$a_1$}; \node at (1.25,6.25) {$a_2$}; \node at (1.75,6.25) {$a_3$}; \node at (2.25,6.25) {$b_4$};
    \node at (0.75,5.75) {$a_1$}; \node at (1.25,5.75) {$c_2$}; \node[fill=babyblue] at (1.75,5.75) {$a_3$}; \node at (2.25,5.75) {$b_4$};
    \node at (0.75,5.25) {$a_1$}; \node at (1.25,5.25) {$b_2$}; \node at (1.75,5.25) {$a_3$}; \node[fill=babyblue] at (2.25,5.25) {$b_4$};
    \node[fill=babyblue] at (0.75,4.75) {$b_1$}; \node at (1.25,4.75) {$a_2$}; \node at (1.75,4.75) {$b_3$}; \node at (2.25,4.75) {$b_4$};
    
    \draw[->, thick] (2.8, 5.75) -- (3.3, 5.75);

    \draw[fill=white] (3.5,6) rectangle (4,6.5) node[pos=.5] {A};
    \draw[fill=white] (3.5,5.5) rectangle (4,6) node[pos=.5] {B};
    \draw[fill=white] (3.5,5) rectangle (4,5.5) node[pos=.5] {C};
    \draw[fill=white] (3.5,4.5) rectangle (4,5) node[pos=.5] {D};

    \foreach \y in {4.5, 5, 5.5, 6} {
        \foreach \x in {4, 4.5, 5} {
            \draw (\x,\y) rectangle (\x+1,\y+0.5);
        }
    }
    
    \node at (4.25,6.25) {$a_1$}; \node at (4.75,6.25) {$a_2$}; \node at (5.25,6.25) {$a_3$}; \node at (5.75,6.25) {$b_4$};
    \node at (4.25,5.75) {$a_1$}; \node at (4.75,5.75) {$b_2$}; \node[fill=green] at (5.25,5.75) {$c_3$}; \node at (5.75,5.75) {$b_4$};
    \node at (4.25,5.25) {$a_1$}; \node at (4.75,5.25) {$c_2$}; \node at (5.25,5.25) {$a_3$}; \node[fill=green] at (5.75,5.25) {$a_4$};
    \node[fill=green] at (4.25,4.75) {$c_1$}; \node at (4.75,4.75) {$a_2$}; \node at (5.25,4.75) {$b_3$}; \node at (5.75,4.75) {$b_4$};

    \end{tikzpicture}
    \caption{Example of mutation: three of the four individuals have a single gene selected for mutation. Adapted from ~\cite{Meloni2016}.}
    \label{figure:mutationmechanism}
\end{figure}
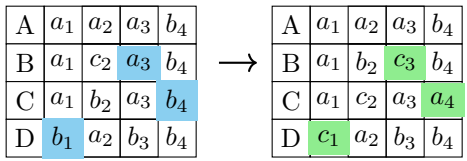
With a total number of four genes per individual, a single gene per individual is selected for mutation, and at least three individuals per generation are mutated to enable the exploration of parameter space and to ensure the convergence of the algorithm. If more than three duplicates are present after the gene pool crossover, more mutations are performed. For the mutation, as shown in \cref{figure:mutationmechanism}, one of the four genes is randomly selected, and the parameter is re-set randomly within the possible values.  

\subsubsection{Evaluation of individuals\label{sec:evaluation}}
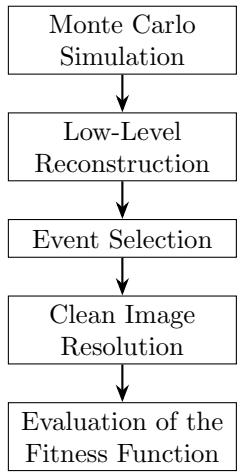
\begin{figure}[!htb]
\centering
\begin{tikzpicture}[
    node distance=5mm and 5mm,
    box/.style={draw, minimum width=3cm, minimum height=5mm, align=center},
    arrow/.style={-Stealth, thick}
]

\node[box] (mc) {Monte Carlo\\ Simulation};

\node[box, below=of mc] (low_level) {Low-Level\\ Reconstruction};
\node[box, below=of low_level] (event_selection) {Event Selection};
\node[box, below=of event_selection] (image_reconstruction) {Image Reonstruction};
\node[box, below=of image_reconstruction] (image_resolution) {Clean Image\\ Resolution};
\node[box, below=of image_resolution] (fitness) {Evaluation of the\\ Fitness Function};


\draw[arrow] (low_level) -- (event_selection);
\draw[arrow] (event_selection) -- (image_reconstruction);
\draw[arrow] (image_reconstruction) -- (image_resolution);
\draw[arrow] (image_resolution) -- (fitness);
\draw[arrow] (mc) -- (low_level);

\end{tikzpicture}
\caption{Steps needed to evaluate an individual in the course of \gls{ga}. The multi-stage Monte Carlo simulation is followed by low-level reconstruction, event selection, image reconstruction, derivation of the clean image resolution, and the evaluation of the fitness of the individual.} 
\label{figure:stepsperindu}
\end{figure}

Evaluation of an individual comprised several steps illustrated in \cref{figure:stepsperindu}.

\begin{figure}[!htb]
\begin{center}
\includegraphics[width=0.98\textwidth]{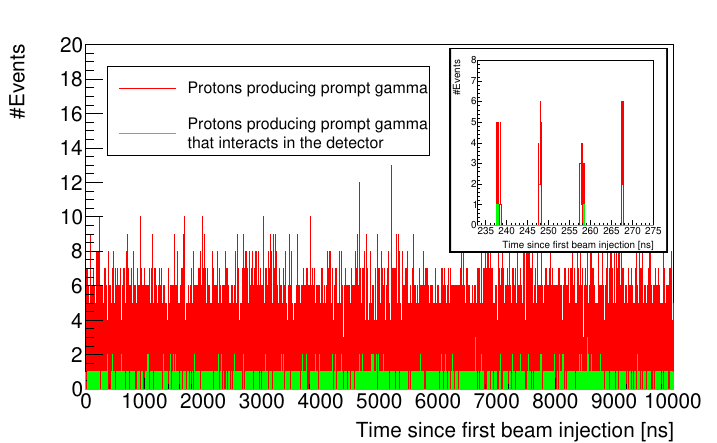}
\end{center}
\caption{Time ion of the simulated \glspl{PG} reflecting the time structure of the clinical proton beam. Red: all events leading to a \gls{PG} production; green: a subset leading to an interaction of \gls{PG} in the detector, example for SL=10, AL=30, SSD=\SI{200}{\mm} and SAD=\SI{200}{\mm}.}
\label{figure:TimingStructure}
\end{figure}

A multi-stage SiFi-CC Compton camera simulation framework was created based on the Geant4 package~\cite{Geant4} in version 10.4.p03 with a physics list QGSP\_BIC\_HP\_EMZ~\cite{PhysicsListGuide}. Such a choice was shown to reproduce fairly well the experimental gamma spectrum \cite{WronskaKasper2021}. A proton beam of energy \SI{130}{MeV} with characteristics similar to a cyclotron beam in \gls{pbs} is shot onto a phantom target. For each individual, \num{4E9} protons are simulated which corresponds to about \numrange{10}{40} \gls{pbs} distal spots and ensures a sufficiently small statistical uncertainty of the derived components of the fitness function that are evaluated based on these statistics (i.e. all but the clean image resolution $\sigma_\mathrm{CleanImg.}$). Proton time distribution assumes a cyclotron cycle time of \SI{9.9}{\nano\s} and an extraction time of \SI{0.8}{\nano\s}. The total number of protons per bunch of 100 was assumed. Within a single bunch, the distribution of time intervals between the arrival of individual protons was Poissonian. The resulting time distribution of the produced \glspl{PG} is presented in \cref{figure:TimingStructure}.
Interaction of \glspl{PG} produced in the \gls{PMMA} phantom with the Compton camera is modelled in detail, including the generation of optical photons, their propagation in the fibres taking into account realistic optical fibre properties, and their registration in \glspl{sipm}, whose properties like \gls{pde} are also accounted for. This simulation of the fibre behaviour was extensively validated against single fibre measurements. Subsequently, hits triggered within a trigger window $t_W$ of \SI{5}{ns} are grouped into events. Importantly, a single event may include signals originating from different gammas produced by the same proton, or even gammas produced by two different beam protons. This phenomenon is called event mixing.
Additionally, due to the time response of LYSO:Ce \cite{LYSODATA}, the signal integration window is set to \SI{400}{ns} and includes the possibility of pile-up occurrence in triggered fibres.

As part of \gls{llr}, each fibre hit can be reconstructed to recover gamma hit position and energy deposit. The $x$ and $z$ positions are constrained by the hit fibre position, while the $y$ hit coordinate is reconstructed from the ratio of signals registered on the two fibre ends. For this, a calibration function is derived from a separate simulation of a small fibre bundle yielding also uncertainties for all reconstructed parameters. A similar approach is used for the reconstruction of energy deposits, although here a geometric mean of the two signals is used instead of the ratio. In the next step, all fibre hits reconstructed this way undergo clustering. The introduction of this step was driven by the observation that in the simulated events in which a high-energy electron appears as a secondary, often several spatially adjacent fibres respond~\cite{Kasper2020}. From this cluster of adjacent fibre hits we assign the event trigger time as given by the earliest time among the hits, while the energy is the sum of energy deposits of the contributing fibre hits. The coordinate of the cluster position, along the detector axis, is given by the position of the first fibre layer within the cluster. In the other two dimensions, the positions are built by the energy-weighted means of all coordinates in this direction of the contributing fibre hits. This method of determining cluster parameters has been tested on Monte Carlo simulations and gave better results than other tested methods.

Afterwards, each event consists of a list of clusters. To be considered a signal event (a so-called distributed Compton event), it must fulfil the following conditions which we take from Monte Carlo information:
\begin{itemize}
    \item the event was produced by a single gamma (no random coincidence or pile-up),
    \item the first interaction of this gamma is a Compton effect,
    \item the Compton electron interacted in one of the detector modules,
    \item the Compton-scattered gamma interacted in the detector again, in the other module.
\end{itemize}
All other events are denoted as background events. This defines the parameters $N_\mathrm{Dist.Comp.}$ and $N_\mathrm{Back.}$ for the fitness function.

The event selection (\gls{es}) process is based on the reconstructed clusters. To be selected as a Compton event, the event has to fulfil the following cuts:
\begin{itemize}
    \item the event contains only one cluster in the scatterer,
    \item the electron energy should respect Compton kinematics assuming full energy deposition of the original gamma in the two detector modules, 
    \item a variation of a distance of closest approach filter \cite{MacKin2013} is used. A Compton cone is calculated and its distance to the expected beam axis is determined. The distance is not allowed to exceed \SI{20}{mm}. Any cone intersecting the beam axis is accepted.
\end{itemize}
Here, the energy and position assigned to the cluster in the scatterer are considered to be the energy and position of the Compton electron. The position of the cluster with the highest energy is taken as a position for the interaction of the scattered gamma and the sum of the energy of all clusters in the absorber is used as the energy of the scattered gamma.

An event reconstructed this way is considered correctly selected Compton event (and such being counted in $N_ \mathrm{Sel.Dist.Comp.}$) if the underlying Monte Carlo event is labelled as distributed Compton event and the reconstructed energies and positions of electron and scattered gamma match the Monte Carlo ones within the expected resolutions derived from clusters with an energy of \SI{0.511}{\mev} relaxed by a factor of~$2\sigma$. Hence, the reconstructed positions and energies should lie within these margins, \cref{eq:third}.

\begin{align}
\begin{split}
    \left| y_{\text{reco}} - y_{\text{MC}} \right| &< 30 \, \text{mm}, \\
    \left| x_{\text{reco}} - x_{\text{MC}} \right| &< 2 \times \text{pixel pitch}, \quad \text{same for } z,  \\
    \left| dE_{\text{reco}} - dE_{\text{MC}} \right| &< 0.12 \times dE_{\text{reco}}, \label{eq:third}
    \end{split}
\end{align}
where the pixel pitch is \SI{1}{mm}. All events selected as Compton events but not fulfilling these criteria constitute $N_\mathrm{Sel.Back.}$.

\begin{figure}[htbp]
\begin{center}
\includegraphics[width=0.98\textwidth]{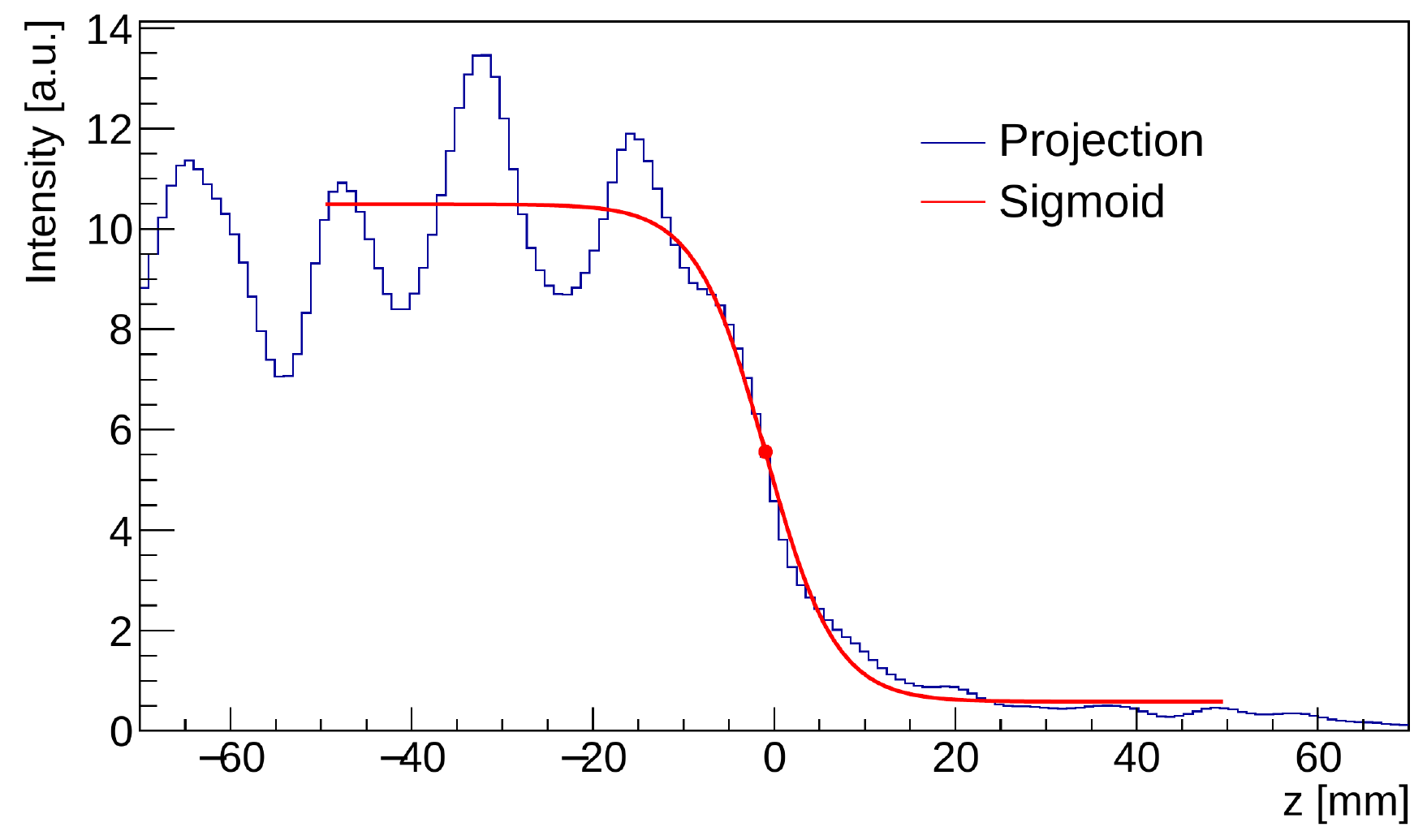}
\end{center}
\caption{The distal fall-off position in a PG depth profile as the inflection point (red dot) of the fitted sigmoid function.} 
\label{figure:DistalFallOffDet}
\end{figure}

The clean image resolution, $\sigma_\mathrm{CleanImg.}$ is obtained by considering 50 subsets of 300 correctly selected distributed Compton events for image reconstruction. Enough variability over the subsets is necessary, so at least 1500 correctly selected events are required in the complete data set. If event selection from \num{4E9} protons fails to yield enough correctly selected events, an additional \num{1E9} protons are simulated. If statistics remain insufficient, the setup is assigned a fitness of 0. 
Otherwise, image reconstruction is performed using a 2D implementation of the iterative \gls{mlem} algorithm \cite{Wilderman1998}. Each reconstructed image is post-processed with a \SI{2}{mm}-kernel Gaussian filter and projected into a 1D profile. The distal fall-off position in each reconstructed image is determined as the inflection point of a sigmoid fit
(see \ref{figure:DistalFallOffDet}). 
The clean image resolution is then determined by the root mean square of the mean distal fall-off position among the profiles reconstructed for each subset.

\subsection{Implementation \label{sec:implementation}}

The processing of \gls{ga} has been fully automated with the help of an SQL (Structured Query Language) database storing the settings and fitness values of the individuals, the necessary directory and executable paths, and the status of the simulation and data analysis process for all individuals. The first-generation individuals are generated randomly and stored in the SQL database. For each individual, all steps described in \cref{figure:stepsperindu} are executed in separate jobs on a batch system. The simulation status is tracked by updating the SQL database and jobs can automatically be restarted if needed.
Once all individuals within a generation have been evaluated, the generation’s fitness is calculated, followed by the application of selection, crossover, and mutation operations. The newly created individuals for the subsequent generation are saved in the database, initiating the processing for this new generation. The framework will automatically halt the algorithm upon meeting the convergence criteria or after evaluating the tenth generation.

\subsection{Capability of range shift detection}
For the best-performing setup, an additional simulation of \num{4e9} protons for a Bragg peak position shifted by \SI{5}{mm} was performed to investigate the range-shift retrieval capability of the setup.
After \gls{llr} and \gls{es}, the simulation data were used to reconstruct images and, in the last step, to derive a resolution of the distal fall-off position. The goal of using a Compton camera in range verification for proton therapy should be to verify the beam range at an individual \gls{pbs} spot level. However, verification utilising data from multiple spots would also be highly advantageous. The data sets for the simulated Bragg peaks correspond to \num{40} \gls{pbs} spots, assuming \num{1e8} impinging protons per beam spot. Image reconstruction on bootstrapped data sets is used here, allowing to derive the dependence of the resolution on the input event statistics. Event subsets of \num{1e8}, \num{2e8}, \num{5e8}, \num{1e9} have been evaluated in two versions: using the Monte Carlo truth information of the registered events, and using experiment-like data, i.e. resulting from \gls{llr}+\gls{es}. The first constitutes a reference showing what could be achieved with a perfect detector resolution and a perfect event selection. In both cases, all selected events were used. In total, \num{100} subsets were assessed for each investigated statistics. For each subset, image reconstruction was performed, and the resulting distal fall-off position was determined. The standard deviation of the distal fall-off positions determined for a given subset size was used as a measure of image resolution in this respect.

\section{Results}
\subsection{Optimisation results}
\begin{figure}[!htb]
\begin{center}
\includegraphics[width=0.98\textwidth]{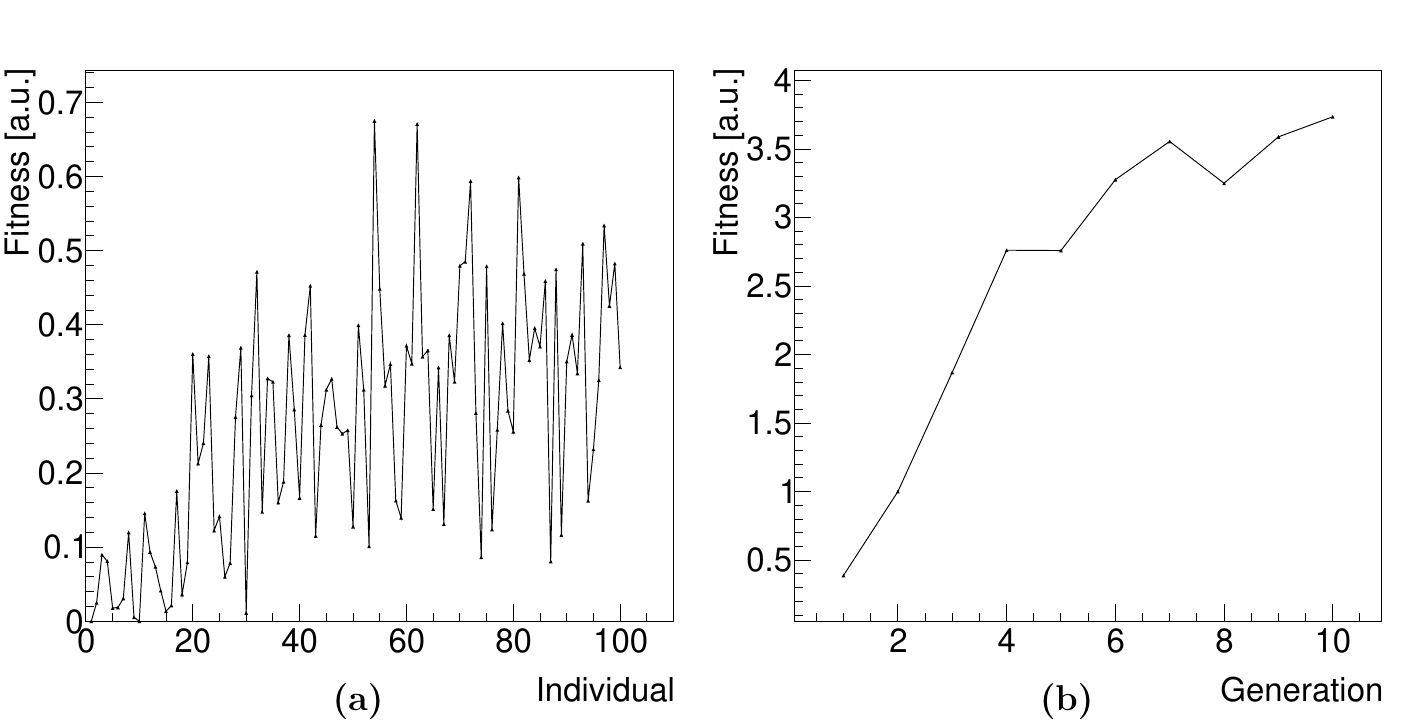}
\end{center}
\caption{(a) Fitness of the individuals evaluated by \gls{ga}, intervals of 10 on the horizontal axis correspond to generations; (b) summed fitness values of whole generations.} 
\label{figure:GaResults}
\end{figure}

\Gls{ga} terminated at the maximum number of ten iterations, since the convergence condition of \cref{algo:ga} was not met before. The fitness values for the 100 assessed individuals are presented in \cref{figure:GaResults}(a). \cref{figure:GaResults}(b) shows the fitness values summed over each generation.
The upward trend in generation fitness, shown in \cref{figure:GaResults}(b), indicates that the algorithm is approaching convergence, with saturation beginning around generation~6. The same trend is also visible in the evolution of the fitness of individuals, shown in \cref{figure:GaResults} (a), with more fluctuations. Some deviations are likely the result of mutations or unfavourable crossover of genes. Individuals yielding very high fitness compared to other individuals in the same generation are likely to be transferred unchanged to the next generation due to the applied principle of elitism, and re-evaluated. Consistently repeating high fitness value means it results from a favourable selection or mutation; otherwise, it was rather caused by a statistical fluctuation.

\begin{figure}[!htb]
\begin{center}
\includegraphics[width=0.98\textwidth]{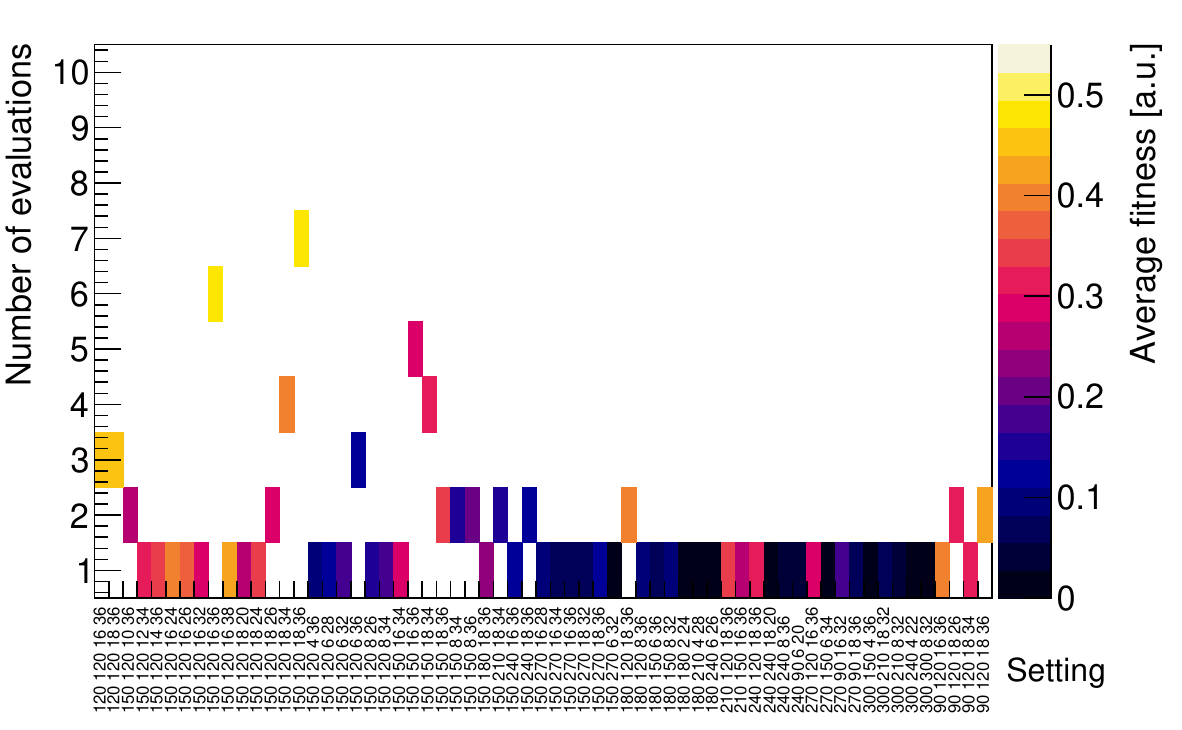}
\end{center}
\caption{Number of evaluations and  average fitness for tested individuals, represented by their parameter settings in the form \gls{ssd}~\gls{sad}~\gls{sl}~\gls{al}.} 
\label{figure:SetsvsInd}
\end{figure}

Multiple evaluations of the same individual can be used to estimate the statistical uncertainty of the fitness of individuals. To achieve this, the sample variance is calculated for each individual assessed a minimum of three times. As seen in \cref{figure:SetsvsInd}, eight distinct individuals are evaluated at least three times. The variances of their fitness values fall within the \qtyrange{9}{25}{\percent} range, leading to an average uncertainty on the fitness of \SI{15}{\percent}. This shows that the chosen convergence criterion is quite stringent, confirming that the algorithm has successfully converged.

In addition, multiple evaluations can mitigate the impact of statistical variations of the outcomes by taking the mean fitness for each individual in the end. Those mean fitness values, along with the number of evaluations, are shown in \cref{figure:SetsvsInd}. The five best-performing setups are listed in \cref{table:garestable}.

\Cref{figure:SetsvsInd} illustrates that throughout the \gls{ga} process, \num{64} distinct sets of optimisation parameters were assessed. This, combined with the observation that the top solutions mainly vary by a single step size in one parameter, indicates that the algorithm began converging to an optimum fairly early. This aligns with the pattern observed in \cref{figure:GaResults}.

\begin{table}
\begin{center}
\begin{tabular}{l|l|l|l|l|l}

Fitness&Number of evaluations &\gls{ssd}&\gls{sad}&\gls{sl}&\gls{al}\\
\hline
\textbf{0.493}&6&\textbf{150}&\textbf{120}&\textbf{16}&\textbf{36}\\
\hline
0.471&7&150&120&18&36\\
\hline
0.455&3&120&120&16&36\\
\hline
0.451&3&120&120&18&36\\
\hline
0.426&1&150&120&16&38\\
\end{tabular}
\caption{The set of best-performing individuals based on their average fitness, the fittest one is bolded.}
\label{table:garestable}
\end{center}
\end{table}

The individuals presented in \cref{table:garestable} exhibit only minor variations in their settings. Given the uncertainty of the determined fitness values, the outcomes for the top five individuals are similar. Nonetheless, this group of individuals significantly outperforms the other evaluated settings. Differences among individuals within this group cannot be assessed with the provided statistics. However, the evident convergence of the \gls{ga} supports the choice of statistics used in the simulations to assess each individual, even though it results in an uncertainty of \SI{15}{\percent}. With these statistics, the \gls{ga} accumulated a total of $\sim$\num{1.2E6} CPU hours. Increasing the number of simulated events would have further extended the runtime, rendering the process unfeasible.

Following the findings in \cref{table:garestable}, the parameter settings showing the best average fitness are interpreted as the best solution found by the \gls{ga} (first row in \cref{table:garestable}) although this was not the best-performing individual in the last generation.

\subsection{Data rate consideration}
In \cref{table:LLRBestGeo}, representative numbers of signal and background coincident events in the optimised setup for a simulated Bragg peak position of \SI{0}{\mm} (centrally in front of the Compton camera) are collected. The maximum data rate that \gls{fee} and \gls{daq} would need to handle depends on the number of recorded coincidences and the acquisition time. Delivery of a single \gls{pbs} spot of \num{1E8} protons takes \SI{9.9}{\ms} which translates to a data rate of 1.7~Mevents/s when only saving coincident events.
\begin{table}[!htb]
\begin{center}
\begin{tabular}{lc}
Event          & Number of  recorded events\\
type               & per \num{4e9} protons $/ 10^3$ \\
\hline
All coincidences   & 674.3(8) \\
Compton signal & 119.9(3) \\
Background     & 554.5(7)                                           
\end{tabular}
\caption{Numbers of recorded coincident events produced with \num{4E9} protons simulated for a Bragg peak position of \SI{0}{\mm}.
\label{table:LLRBestGeo}}
\end{center}
\end{table}

For each triggered coincidence, an estimated $2\times3\times N_\mathrm{Fibres}$ data packets need to be sent through the data stream consisting of channel ID, number of registered optical photons (signal integral), and the time stamp of the triggered \glspl{sipm}, where the mean fibre multiplicity per event for the fittest setup is $N_\mathrm{Fibres}=5.7$.

Moreover, the coincident events constitute only \SI{11}{\percent} of the whole expected recorded event sample. The remaining \SI{89}{\percent} are events in which only one module responded, that cannot be used for image reconstruction. Thus, the ability to trigger selectively on coincidences at the hardware level is crucial to remove those data from the data stream.

Based on the event statistics obtained for the best-performing geometry, the imaging sensitivity of such a Compton camera, defined as the ratio between the number of events useful for \gls{ir} and the number of impinging protons, yields \num{5.58(1)e-5}. This number accounts for the fact that the event selection qualifies only \SI{33}{\percent} of all coincident events to image reconstruction.

\subsection{Capability of range shift detection}

\begin{figure}[!htb]
\begin{center}
\includegraphics[width=0.48\textwidth]{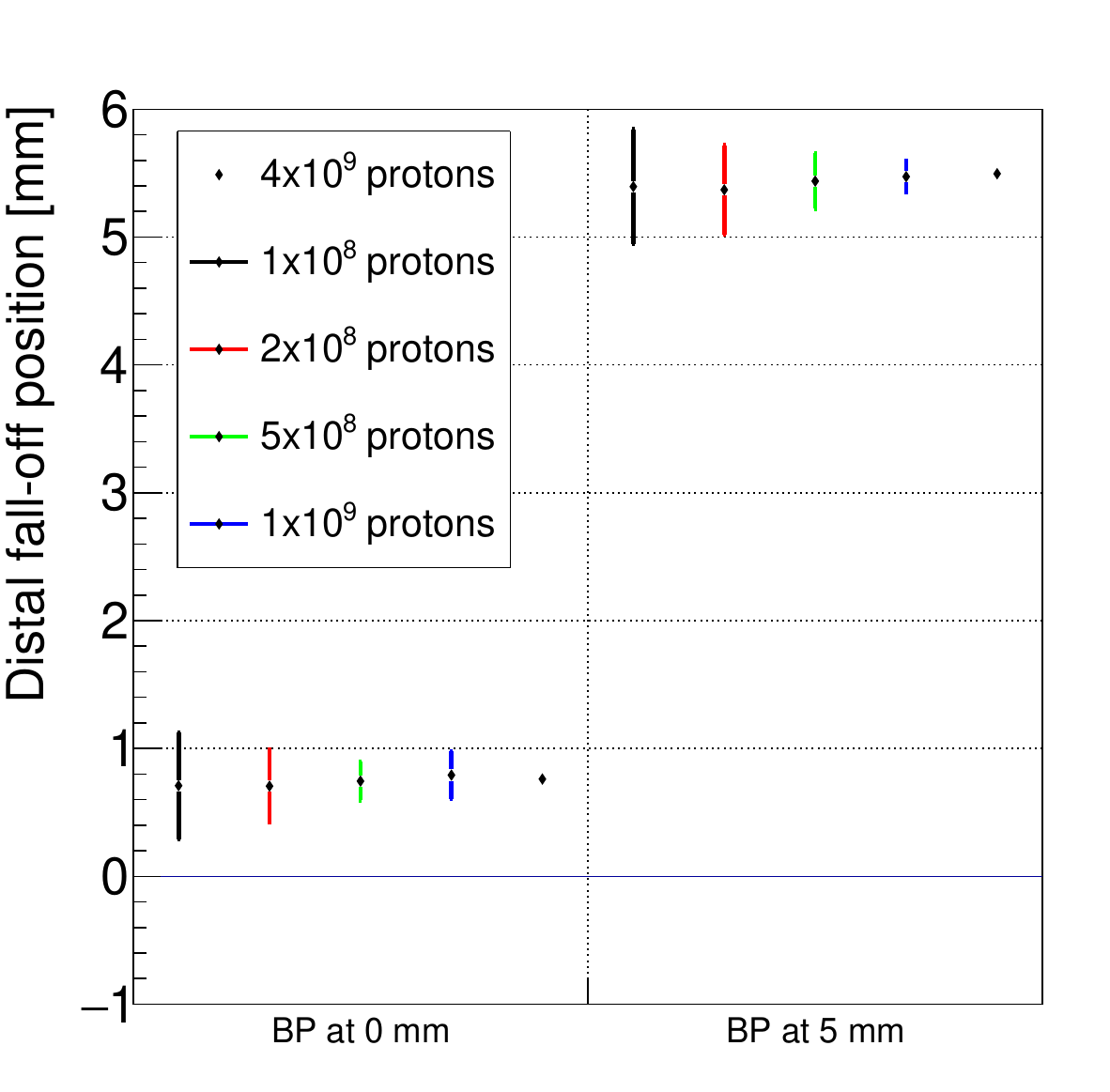}
\includegraphics[width=0.48\textwidth]{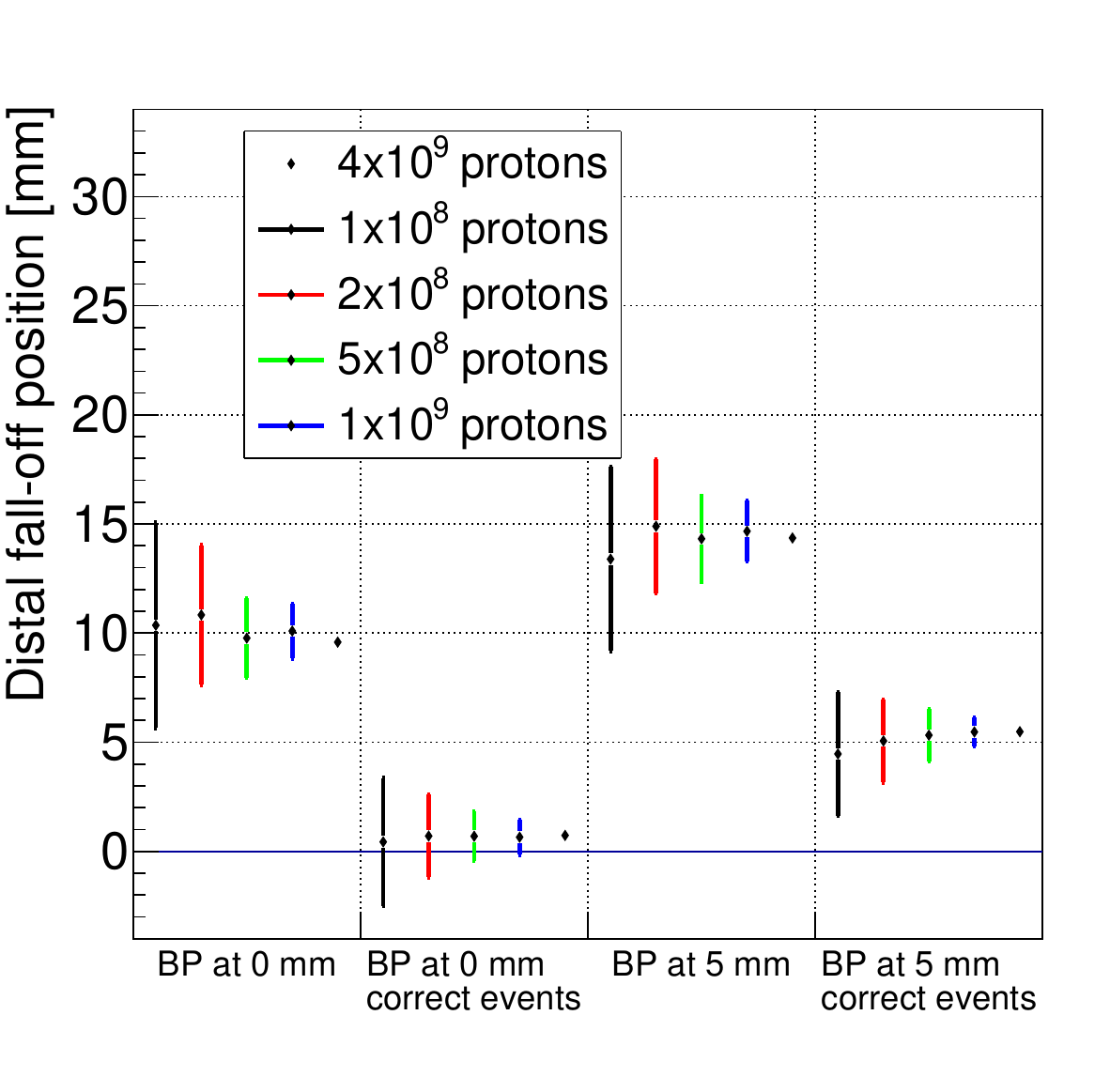}
\hspace{0.12\textwidth}\textbf{(a)}
\hspace{0.45\textwidth}\textbf{(b)}
\end{center}
\caption{Study of distal fall-off (inflection point of a sigmoid fit) position resolution as a function of statistics used for image reconstruction (see legend). Data from the simulation of the best-performing setup in the \gls{ga}. (a) Results based on image reconstruction with Monte Carlo truth data. (b) Results with reconstructed data filtered through event selection and for correctly selected events are also shown.} 
\label{figure:MCExpShift}
\end{figure}

Simulation of a single \gls{pbs} spot of \num{1E8} impinging protons leads to \num{2560} distributed Compton events, whose Monte Carlo truth information can be fed to image reconstruction delivering a reference figure. The results of the study of statistical precision in the distal fall-off determination are shown in \cref{figure:MCExpShift}(a). The samples of different numbers of impinged protons are represented in different colours. Additionally, the result obtained from the analysis of the full data set of \num{4e9} protons is shown for reference. Due to the inability to form subsets for this case, error bars are not included for that data point. However, the resolution can be inferred by extrapolating the observed trend from the other data points.
The results for the Monte Carlo truth data show the expected behaviour. The resolution improves with the increasing statistic used in the subsets. The determined distal fall-offs for both Bragg peak positions are consistent within the tested subset statistics. The values of these positions do not fit the simulated positions of the Bragg peak, but instead show a constant submillimeter shift. Since the spatial distribution of prompt gamma emission vertices is not the same as that of the deposited dose, such a shift is expected. Important for the ability of a detector to detect a range shift in the proton beam is that the achieved resolutions are sufficiently small compared to the range shift. The detected range shift amounts to \SI{4.7}{\mm}, and the resolution  for a single \gls{pbs} spot is \SI{0.4}{\mm}; hence,  the \SI{5}{\mm} shift can be clearly detected.

\Cref{figure:MCExpShift}(b) shows results for reconstructed data filtered through event selection data, again in two versions: with a perfect selection, i.e. exploiting only correctly identified Compton events, and with a realistic selection, including all surviving background. The presence of background is a source of two effects: a change of the constant offset between the distal fall-off position and the Bragg peak position, and a deterioration of statistical precision. In contrast to the second issue, the first issue does not affect the setup capability to determine range deviations. For the data set that includes incorrectly selected events, the determined range shift is \SI{4.8}{\mm}. However, the resolution achieved for a single $10^8$-proton spot is of the order of \SI{5}{\mm} only. Increasing the statistics brings the expected improvement: for five \gls{pbs} spots the resolution is as good as \SI{2}{\mm}. Meeting more ambitious goals requires either a higher efficiency, better event selection methods, or a combination of both.

\section{Discussion}

\subsection{Performance of the optimised setup}

The evaluation of the best solution given by the \gls{ga} shows that a setup of the \gls{sificc} with \num{16} layers in the scatterer, \num{36} layers in the absorber, a distance between the scatterer and the source position of \SI{150}{\mm} and a distance between the two modules of \SI{120}{\mm} is capable of detecting a beam range shift based on \num{5e8} impinging protons with a \SI{2}{\milli\metre} precision.

The performance study is based on simulation data generated with a multi-stage simulation framework, which takes into account many properties of the proton therapy environment and the detection setup. However, interactions of secondary particles within the target have not been included. These interactions strongly depend on the target characteristics, such as the tumour type and its location in the patient. Limiting the analysis to events where gamma rays with energies above \SI{1}{\mev} are detected could help mitigate this issue, although it further decreases detection efficiency.

The work also neglects the presence of other secondary particles of the proton beam, mostly neutrons with a continuous energy spectrum~\cite{Ortega2015}. In the secondary phase space obtained in the course of simulations for this work, a ratio between prompt gammas and neutrons of \SI{1.1} has been observed, which fits the findings reported in~\cite{Ortega2015}. The influence of this neutron background on the recorded coincidences will be investigated. Although neutron-induced background in the coincident event sample is not expected to be very large ($\approx$\SI{20}{\percent} based on~\cite{Ortega2015}), real gamma-induced coincidences may be contaminated with additional neutron-induced hits, reducing the number of usable Compton events. A possible remedy to the effect of additional secondary particles may be the use of a better \gls{es}. Recently, promising approaches based on the application of neural networks have been proposed~\cite{Munoz2021, Majid2022, Lerendegui2022,York2021,Polf2022,Barajas2023}. Such approaches are also under development for the \gls{sificc} project.

Data acquisition rate capability of about $10^6$~events/s needed for the proposed setup can be reached by several systems discussed in~\cite{Wong_2024}.

Another component of the data analysis sequence that holds significant promise for enhancing the resolution of distal fall-off position determination is \gls{ir}. Here, a basic \gls{mlem} algorithm is employed, but it has been shown that more sophisticated techniques can enhance image quality~\cite{Munoz2021,Kohlhase2020}.

However, it should be stressed that the setup with the geometry derived with the \gls{ga}, combined with rather simple \gls{es} and \gls{ir} algorithms, already provides the capability to detect range shifts with the precision relevant for clinical applications. Given the above-mentioned possibilities of improvement, the goal of range verification based on the statistics of a single \gls{pbs} spot seems to be within reach. The reported range shift sensitivity is comparable to other measurements and simulation studies for Compton cameras reported in the literature~\cite{Munoz2021,Fontana2019,Draeger2018}.

\subsection{Feasibility of the optimisation procedure}

As indicated above, the classical approach to solving an optimisation problem via a full scan of the parameter space was infeasible due to the enormously large CPU runtime needed.  The use of \gls{ga} allowed the number of evaluated potential candidate geometries to be reduced from the \num{7040} required for a classical parameter scan to \num{100} (though physically some of those \num{100} geometries represented the same setup geometry). The simulation and evaluation process of a single individual took between \num{8000} and \num{14000} CPU hours, depending on the chosen geometry. The entire \gls{ga} search needed about \num{1.2e6}~CPU hours and thus was only feasible on a computing cluster.

An alternative approach, as has been successfully demonstrated for single-module cameras with passive collimation~\cite{Huisman2023}, could rely on building an analytical model based solely on the geometrical and material properties of the setup. A full analytical model in the sense of a system matrix is often used in CT, SPECT or PET. However, in the case of a Compton camera, such an approach would not be feasible, since the measurement space is quasi-continuous and multi-dimensional. Instead, we work with list-mode data and calculate system matrix elements on the fly, or pre-compute only those elements of the system matrix that are needed. The quality of image reconstruction relies both on the statistics of the registered coincident events, and the quality of reconstruction of Compton cones, which is defined by energy- and position resolutions of the modules, whereby the two are mutually dependent. Moreover, a modified acceptance of the Compton camera when changing its geometrical parameters also leads to a change in the sample of event topologies that can be registered with it. This may lead to a change in the mean quality of reconstruction of the Compton cones, and consequently the whole image. The problem is much more complex than in passively collimated setups, making this approach unsuitable for Compton cameras.

\section{Conclusion}

The \gls{sificc} detector employs a Compton camera design aimed for verifying the beam range in proton therapy. It utilises LYSO:Ce fibres with a dual \gls{sipm} readout arranged in layers and stacks, forming two modules. The study aims to maximise its ability to detect shifts in the proton beam range under clinical conditions.

A multi-stage simulation framework was developed to simulate the \gls{sificc} candidate configurations in a CPU-economic way in clinical conditions. It contains a complete interaction chain from a proton beam hitting a target to the interaction of the optical photons in the \glspl{sipm}. A complex data processing scheme was proposed, starting from low-level reconstruction, event selection, image reconstruction and finally the analysis of the statistics-dependent resolution in range shift retrieval.

The \gls{ga} optimised the number of layers per module and the distances between the scatterer and source as well as between the scatterer and the absorber. It ended at ten generations, each consisting of ten individuals, and the outcome shows the onset of convergence, even if the rather strict imposed convergence condition was not met. The best performance was found for the setup with 16 fibre layers in the scatterer, 36 layers in the absorber, a distance between the scatterer and the source of \SI{150}{\mm} and a distance between the scatterer and the absorber of \SI{120}{\mm}. The optimisation, necessary only once in the construction phase of the setup, was feasible runtime-wise only because of the usage of \gls{ga} and innovations in the simulation and data analysis sequence. 

The best-performing \gls{sificc} setup was able to measure a beam range shift of \SI{5}{\mm} based on the detector response to \gls{PG} radiation originating from the interactions of \num{5e8} impinging protons in the phantom with a \SI{2}{mm} resolution. A rather simple event selection method was used. The resulting imaging sensitivity of such a setup was \num{5.58(1)e-5}.

The findings of this work constitute the next milestone in the  \gls{sificc} detector development.

\section*{Acknowledgements}
This work was supported by the Polish National Science Centre (grant number 2017/26/E/ST2/00618). The authors are grateful to their colleagues from the University of Lübeck, Prof. Magdalena Rafecas and Dr. Jorge Roser, for their careful proofreading of the manuscript,  constructive criticism and helpful suggestions.
\printbibliography
\end{document}